# Lithium and sodium decorated PHE-graphene for high capacity hydrogen storage: A DFT and GCMC study

MA Hongyan, WANG Qing, SUN Huilin, LI Qingyu, WANG Yunhui, YANG Zhihong, ZHAO Huaihong, SHU Huazhong

*Jiangsu Provincial Engineering Research Center of Low-Dimensional Physics and New Energy, College of Science, Nanjing University of Posts and Telecommunications, Nanjing 210023, PR China.*

E-mail addresses: yhwang@njupt.edu.cn; yangzhihong@njupt.edu.cn

## Highlights

- DFT and GCMC simulations reveal the superior hydrogen storage properties of Li/Na-decorated PHE-graphene.
- Li/Na-PHE-graphene achieves exceptional $H_2$ storage capacities of 15.2 wt% (Li) and 12.6 wt% (Na), surpassing DOE targets.
- Charge density difference and electron localization function analyses clarify the mechanism behind enhanced $H_2$ adsorption.
- GCMC results validate DFT predictions, ensuring robust and reliable conclusions.

## GRAPHICAL ABSTRACT

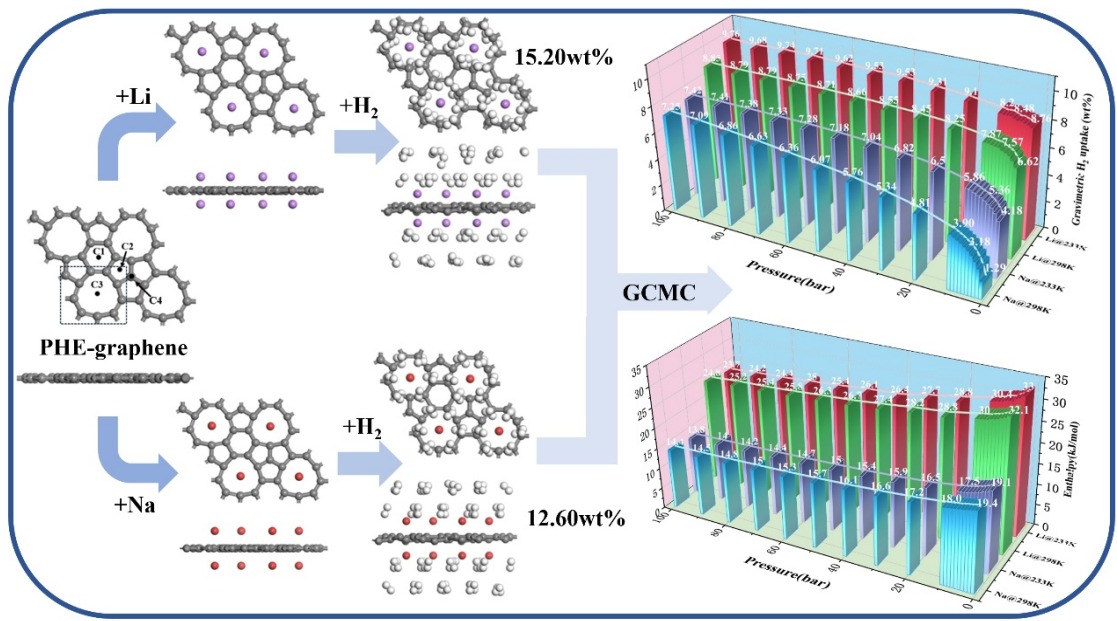

## ABSTRACT

Porous nanocarbon materials are seen as potential excellent materials for hydrogen storage due to their high surface area, excellent cycling stability and favorable kinetics. This study employs Density Functional Theory (DFT) simulations to investigate key property of Li- and Na- modified PHE-graphene, including structural stability, electronic properties, and hydrogen storage capabilities. The results show that when each Li atom adsorbs six hydrogen molecules, the material reaches the maximum hydrogen adsorption gravimetric density of 15.20 wt%. Additionally, through Grand Canonical Monte Carlo (GCMC) simulations, we obtained the hydrogen weight ratios and adsorption enthalpy curves for Li- and Na-modified PHE under varying temperature and pressure conditions. These findings indicate that both Li- and Na-modified PHE-graphene are exceptional candidates for hydrogen storage materials, particularly in mobile applications.

**Keywords:** DFT; Hydrogen storage; PHE-graphene; GCMC simulations

# 1.Introduction

With increasing global dependence on petroleum-based fuels, energy shortages and environmental pollution have garnered significant attention, prompting worldwide efforts to identify clean and sustainable alternative energy resources [1][2]. Hydrogen energy has become a focal point across diverse sectors owing to its wide source, high safety, zero emissions, and superior energy density [3]. Recent progress in hydrogen energy applications has been remarkable, with particularly significant advances in storage technologies. Conventional hydrogen storage approaches - such as compressed gas, liquefied hydrogen, and cryogenic storage systems - generally require either high-pressure containment (typically 350-700 bar) or ultra-low temperature maintenance (below -253°C) [4]. These conventional methods typically demand expensive cryogenic insulation systems and bulky storage infrastructure, significantly increasing both operational costs and safety risks in practical applications [5][6]. In contrast, solid-state nanomaterial-based hydrogen storage systems offer a superior alternative to conventional methods, enabling physisorption mechanisms with enhanced efficiency and practicality [7][8]. Nevertheless, achieving fully reversible hydrogen storage remains a critical challenge for solid-state nanomaterials, particularly in hydrogen-powered vehicles and industrial-scale applications where cycling stability and rapid kinetics are essential [9]. Current technical roadmaps for solid-state hydrogen storage systems specify that hydrogen fuel cell vehicles must achieve a minimum gravimetric capacity of 5.5 wt% by 2025 to meet commercial viability thresholds, as established by DOE and ISO standards [10]. To achieve reversible adsorption of hydrogen by the material the adsorption energy of the materials should range from 0.1 to 0.7 eV [11]. Researchers worldwide are actively investigating novel materials for high-capacity hydrogen storage. Carbon-based nanostructures - including nanotubes [12], graphene [13], fullerenes [14], and graphyne [15] have emerged as particularly promising candidates due to their unique hydrogen adsorption properties. Their structural properties, including low mass, high hydrogen storage capacity, and thermal/chemical stability, enable effective hydrogen storage under diverse conditions. However, most current materials suffer from weak hydrogen adsorption energies (typically < 0.1 eV), leading to insufficient storage capacities that fall short of practical application requirements [16]. Alkali metal doping in carbon-based nanomaterials can significantly enhance their hydrogen adsorption capacity through multiple mechanisms [17]. Therefore, Li, Na, and K atom modifications are frequently applied to boost hydrogen storage capacity [18][19].

Recent studies have demonstrated the potential of metal-modified carbon-based materials for hydrogen storage applications. Using first-principles DFT calculations, Zhang et al. demonstrated that double-side Li-decorated Irida-graphene achieves a 7.06 wt% hydrogen storage capacity, exceeding the DOE 2025 target for onboard hydrogen storage systems [20]. Kim et al. studied Ca-modified Polygon-Graphenes by first principles calculations, discovering that each Ca atom could adsorb 5 and 6 $H_2$, it is anticipated that the $H_2$ gravimetric capacity will attain 6.8wt % for biphenylene [21]. Chen et al. modified Tri with Li, Na, K, and Ca, found that Tri and K- modified Tri did not stably adsorb hydrogen. Na- modified Tri could adsorb 14 hydrogen molecules (8.26wt%), while Li- modified Tri and Ca- modified Tri adsorbed 24 molecules, reaching storage densities of 13.99wt% and 12.77wt%, respectively [22]. Beyond the studies mentioned earlier, other nanostructures modified by metal atoms, such as double-sided lithium modified β12-borophene system [23], sandwich graphene(N) -Sc-graphene (N) structure [24], light metal decorated C20 fullerenes [25], Y-modified two-dimensional porous graphene [26] and lithium-modified C2N [27] are also deemed promising for hydrogen storage applications.

Zeng et al. in their investigation of Li-S batteries first predicted a unique two-dimensional carbon allotrope termed PHE-graphene (PHE), which is a porous nanocarbon materials consisting of 5, 6, and 9 carbon ring structures. PHE offers numerous advantages, such as high specific surface area, excellent dynamic and thermal stability, outstanding mechanical properties, and ultralow lattice thermal conductivity [28]. Researchers are actively evaluating the practical applications of PHE. Jiang et al. systematically investigated transition metal (TM)-decorated PHE-graphene as high-efficiency catalysts for water splitting and achieved breakthrough results [29]. Inspired by these studies, we systematically investigated the hydrogen adsorption properties of PHE using DFT and GCMC simulations, different aspects such as binding energies, average adsorption energies, density of states, charge transfers have been explored. At the same time, we calculated and depicted the isothermal adsorption curves for the hydrogen weight ratio and adsorption enthalpy of Li- and Na-modified PHE at various temperatures and pressure by GCMC simulations. Our study demonstrates that Li and Na- decorated PHE shows promising hydrogen adsorption properties.

## 2. Computational details

The VASP is employed for both geometry relaxation and energy convergence [30][31]. Cutoff energy is set at 500 eV [32]. The GGA and PBE functionals are used in this study [33][34]. To avoid interactions between the carbon layers, the interlayer distance is set to 20 Å. We adopted Grimme's DFT-D3 dispersion correction method. Gamma k-mesh is set to 5 × 5 × 1 [35]. We set the convergence threshold for electronic energy to $10^{-5}$ eV and the force convergence criterion for atomic interactions to $10^{-2}$ eV/Å [36]. Additionally, the GCMC simulations of PHE 's H2 adsorption process was evaluated [37][38]. A total of $5 \times 10^6$ simulation steps were performed in the GCMC simulation to make sure the equilibration of the outcome.

Binding energy is an important physical quantity to measure the stability of metal-decorated PHE. It can be computed using the following equation:

$$E_b = (E_{mM-PHE} - E_{PHE} - mE_M)/m \tag{1}$$

where $E_{mM-PHE}$ is the total energy of PHE modified with m alkali metal atoms, $E_{PHE}$ and $E_M$ are the energies of pure PHE and isolated metal atoms, respectively.

The energy of H₂ average adsorption ($E_{ad}$) and energy of H₂ continuous adsorption ($E_c$) are defined by the following equation:

$$E_{ad} = (E_{nH_2-mM-PHE} - E_{mM-PHE} - nE_{H_2})/n \tag{2}$$

$$E_c = E_{nH_2-mM-PHE} - E_{(n-1)H_2-mM-PHE} - E_{H_2} \tag{3}$$

where $E_{H_2}$ is the total energy of H₂, $E_{nH_2-mM-PHE}$ and $E_{(n-1)H_2-mM-PHE}$ is the total energy of alkali metal modified PHE adsorb $n$ and $(n-1)$ H₂, respectively. $E_c$ is an important criterion for judging whether hydrogen is stably adsorbed, it has been shown that when $E_c$ is less than $-0.1$ eV, the material does not stably capture hydrogen molecules. This can be used to determine whether the hydrogen molecules adsorb stably.

The weight density of hydrogen storage was estimated according to the following formula:

$$H_2(wt\%) = \left[\frac{nM_{H_2}}{nM_{H_2}+M_{host}}\right] \tag{4}$$

where $M_{H_2}$ and $M_{host}$ show the weight of H₂ and the host structure, respectively.

The equation of desorption temperature is expressed as:

$$T_D = E_{ads}/\left[k_B\left(\frac{\Delta S}{R} - \ln p\right)\right] \tag{5}$$

The Charge Density Difference (CDD) is determined using the following formula:

$$\Delta_{\rho 1} = \rho_{M-PHE} - \rho_M - \rho_{PHE} \tag{6}$$

where $\rho_{M-PHE}$, $\rho_M$ and $\rho_{PHE}$ denote the electron densities of the matel-PHE complex, isolated metal, and PHE.

$$\Delta_{\rho 2} = \rho_{H_2-M-PHE} - \rho_{H_2} - \rho_{M-PHE} \tag{7}$$

where $\rho_{H_2-M-PHE}$ and $\rho_{H_2}$ correspond to the electron densities of the $H_2$-adsorbed system and isolated $H_2$ molecule.

The Morse potential parameters for atoms in the Dreiding force field are given by the following equation:

$$U(r_{ij}) = D\{exp[\alpha(1-\frac{r_{ij}}{r_0})] - 2*exp[\frac{\alpha}{2}(1-\frac{r_{ij}}{r_0})]\} \tag{8}$$

where $\alpha$ is the force constant, $D$ is the well depth, $r_{ij}$ is the interaction distance, and $r_0$ is the equilibrium bond distance.

### 3. Results and discussions
### 3.1. Li- and Na- modified PHE

The PHE material was first characterized, and its structure was subsequently optimized to achieve higher hydrogen storage capacity. The unit cell is composed of 10 carbon atoms, forming natural two-dimensional potential energy wells that can trap alkali metal atoms. We constructed a 2 × 2 × 1 PHE supercell and optimized the structure, as depicted in Fig. 1(a), the dashed line shows the unit cell. All atoms in the optimized PHE structure remain coplanar, with a = 10.47 Å, b = 10.47 Å [28]. In the PHE structure, only C-C bonds are observed, and the bond distance is 1.447 Å, 1.514 Å and 1.387 Å. Although the material possesses a high specific surface area, its hydrogen adsorption capacity is limited, as indicated by the PHE-H₂ adsorption energy of -0.08 eV, result in suboptimal performance for hydrogen storage applications. Research suggests that alkali metal decoration can increase the adsorption capacity of hydrogen. Therefore, we choose to use Li and Na atoms to modify PHE. To investigate the optimum adsorption sites of Li and Na atoms on the PHE surface, we used VASP to calculate binding energies at different positions. We placed the alkali metal atoms at different positions and optimized them. The positions of the stable metal atoms after optimization are shown in Fig.1 (a). Which are placed in C1 (center of 6 carbon ring), C2 (center of 5 carbon ring), C3 (center of 9 carbon ring) and C4 (top of C atom). Li and Na's binding energies at different sites calculated from Eq. (1), Li at C1, C2, C3, C4 points are -2.38 eV, -2.33 eV, -2.50 eV, -2.10 eV, Na at C1, C2, C3, C4 were -1.71 eV, -1.64 eV, -1.95 eV, -1.59 eV, respectively. We discerned that the maximum binding energy occurs at the C3 position for both lithium and sodium, with values of -2.50 eV and -1.95 eV. The condensation energy of lithium is 1.60 eV [38][39], while that of Na is 1.13 eV [40]. Therefore, Li and Na atoms can stably bind to the PHE structure without aggregation. Finally, we opted to position Li or Na atom symmetrically above and below the center of the 9-membered carbon rings. After structural optimization every carbon atom in the PHE structure is positioned within a single plan, with eight Li and eight Na atoms arranged on either side of the structure as presented in the Fig.1. The lithium atoms are positioned on average 1.24 Å away from the PHE, whereas sodium atoms are located 1.75 Å on average from the graphene surface.

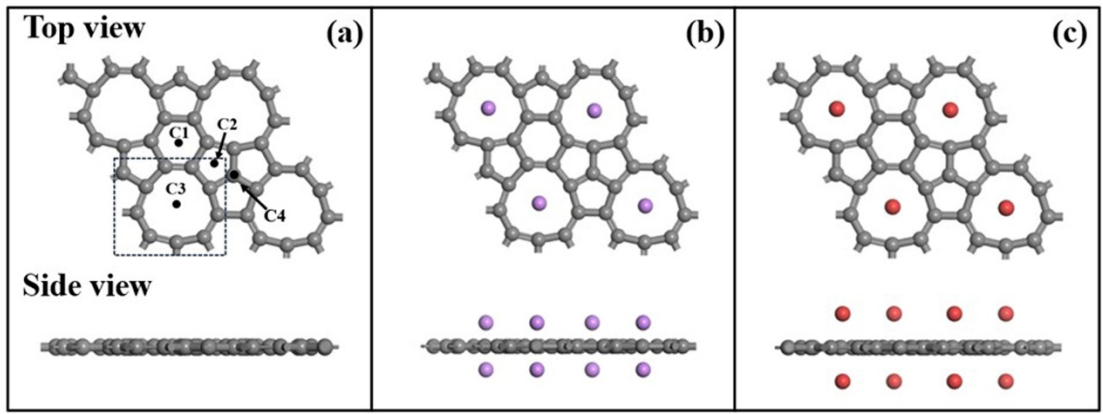

**Fig. 1**. Geometric configurations of optimized (a) PHE, (b) Li- PHE, (c) Na- PHE structures.

In order to examine the charge transfer and bonding characteristics between the PHE monolayer and Li, Na atoms, we compute the Charge Density Difference (CDD) using Eq. (6). As is show in Fig. 2, we can observe that charge redistribution occurring between the metallic atoms and carbon atoms. Charge accumulation occurs between the metallic atoms and the PHE-graphene layer, enhancing the electrostatic field interaction and leading to stronger bonding energies for the metals.

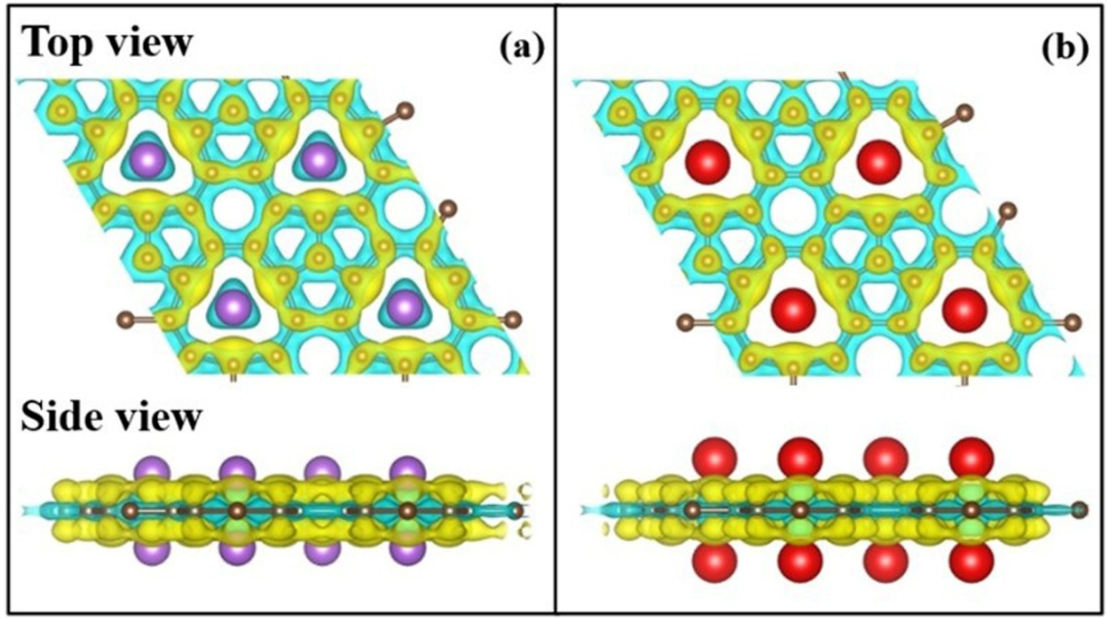

**Fig.2**. The Charge density difference for (a) 8Li- and (b) 8Na-decorated PHE-graphene. Yellow: charge accumulation, cyan: electron depletion.

We use Electron Localization Function (ELF) to study the electronic properties of modified materials. From Fig. 3 we observe that the electrons are mostly clustered around the carbon skeleton. This observation indicates that there is no covalent bond between the metal atom and the carbon skeleton, but an ionic bond is formed. Additionally, some electrons accumulate above the metal atoms. We can predict that placing hydrogen atop a metal atom will be attractive. To enhance the understanding of the ELF, we present the 2D slice diagrams in Fig. 3(c) and (f), illustrating a red area with a value close to 1.0, indicating a high concentration of electrons, and a blue area with a value of zero, indicating the absence of electron distribution. Thus, we observe strong localization between each pair of carbon atoms, demonstrating that they are linked through covalent bonds.

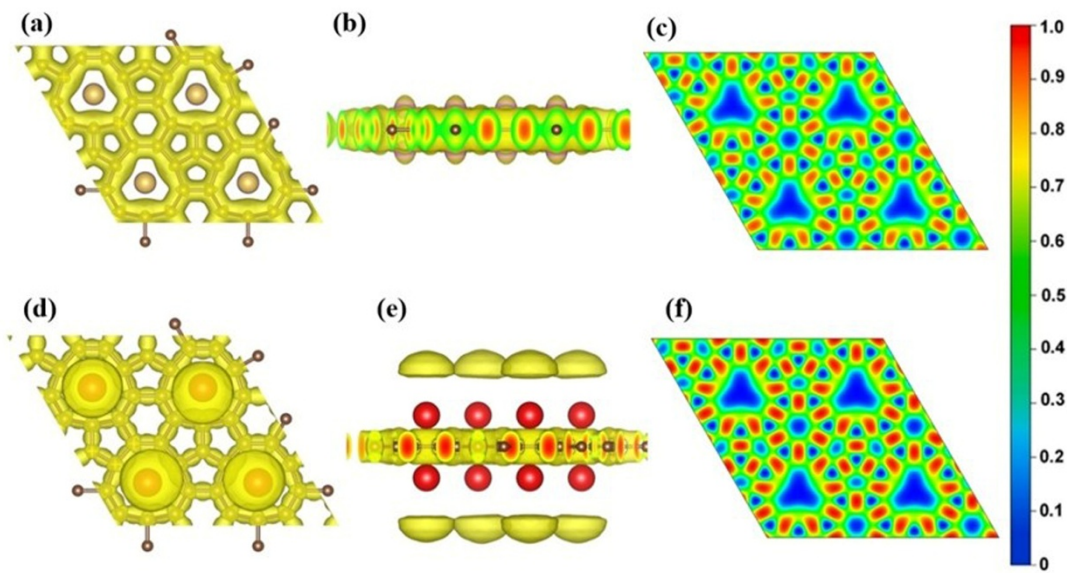

**Fig.3**. The Electron localization function (ELF) of (a)(b)(c) Li- modified PHE-graphene (d)(e)(f) Na- modified PHE-graphene.

**3.2. Li- and Na-modified PHE monolayers adsorbed hydrogen**

Hydrogen adsorption was performed on the Li- and Na- modified PHE structure. Initially, one hydrogen molecule was adsorbed per Li atom, resulting in significant structural deformation due to strong H-Li interactions, as evidenced by an adsorption energy (Ead) of -0.50 eV, shown in Fig. 4(a) and Table 1. Subsequently, the two $H_2$ molecules adsorbed by each Li atom are placed in opposition, and three $H_2$ molecules are placed at the vertices of an equilateral triangle within a nine-membered carbon ring. For the fourth $H_2$ molecule, two configurations were explored: placement above the Li atom or in a quadrangular arrangement. The configuration with the highest adsorption energy was selected, where three $H_2$ molecules maintained a triangular arrangement, and the fourth was positioned above and below the midpoint of a six-membered ring, effectively lying in the same plane shown in Fig. 4(d). The distances between the central $H_2$ molecule and the three nearest Li atoms were 3.391 Å, 3.574 Å, and 3.673 Å, respectively, indicating a triangular Li arrangement favorable for $H_2$ adsorption. For the fifth hydrogen molecule, we opted to position it directly above the lithium atom, based on the configuration in Fig. 4(d), which maintained stability in the optimized structure, as illustrated in Fig.4(e). When adding the sixth atom, considering the triangular stability of the structure, we choose to add three hydrogen molecules in a triangular shape on the basis of Fig.4(c). The post-optimized structure is depicted in Fig. 4(f), featuring hydrogen distributed across two layers, with three hydrogen molecules in each layer, hydrogen storage weight ratio reached 15.20wt%. We measured the distance between the Li and the nearest $H_2$ at 2.038 Å, while the distance to the more distant hydrogen molecule was 4.395 Å. The $E_{ad}$ after the addition of six $H_2$ molecules is 0.20 eV/$H_2$, falling within the optimal adsorption range of 0.20 to 0.40 eV/$H_2$.

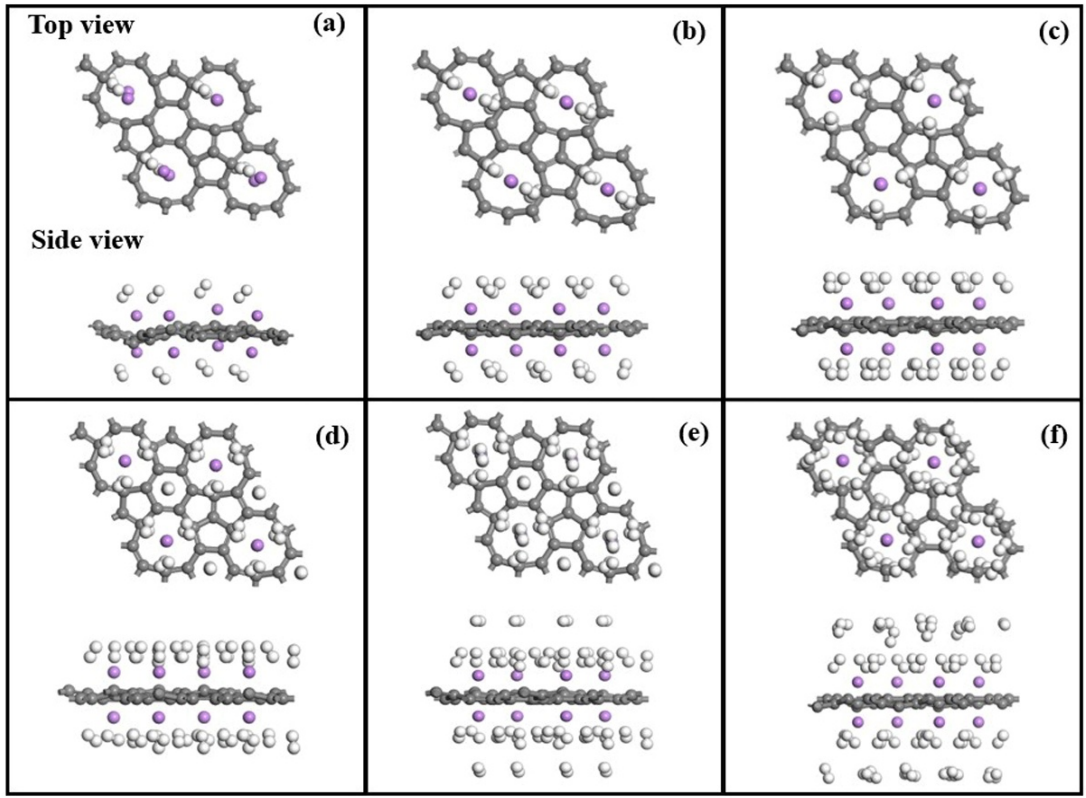

**Fig.4**. The optimized structures of (a) 8$H_2$, (b) 16$H_2$, (c) 24$H_2$, (d) 32$H_2$, (e) 40 $H_2$, (f) 48$H_2$ adsorbed on Li-decorated PHE-graphene.

Next, hydrogen adsorption on Na-modified PHE-graphene was simulated. As shown in Fig. 5 (a)~(f), each Na adsorbed hydrogen increases from 1 to 6. Similar to Li decoration, the first $H_2$ adsorbed beside the metal Na, the carbon structure does not deform this time, which may be because the adsorption energy between Na and hydrogen is not as large as that between Li and hydrogen. Table 1 supports our conjecture that $E_{ad}$ of Na adsorbing a hydrogen is -0.32 eV/$H_2$, lower than Li's $E_{ad}$ of -0.50 eV/$H_2$. When adsorbing four $H_2$ per Na atom, we also utilized the two previously proposed addition methods. By optimizing the structure and selecting the method yielding the highest adsorption energy, we identified the adsorption mode depicted in Fig.5 (d). Different from the adsorption of lithium, the four hydrogen molecules adsorbed by Na form a quadrilateral around the Na atom. Each Na atom can subsequently adsorb 5 or 6 hydrogen molecules, resulting in an optimized structure similar to that of Li-hydrogen adsorption. With six $H_2$ molecules adsorbed per Na atom, the adsorption energy is -0.18 eV per $H_2$, corresponding to a weight ratio of 12.6 wt%.

The process of hydrogen adsorption by Li and Na atom are show in Fig. 4 and Fig. 5. All relevant data are show in Table 1. According to the data calculated by VASP, the $E_{ad}$ of the structure for $H_2$ is calculated by the main material according to Eq. (2), the continuous adsorption energy according to Eq. (3), the weight density of hydrogen storage according to Eq. (4). And the $T_D$ is calculated by Eq. (5) known as the van't Hoff equation [41], indicating that Li- and Na-decorated PHE-graphene can reversibly adsorb hydrogen at temperatures above 255.8 K and 230.2 K, respectively.

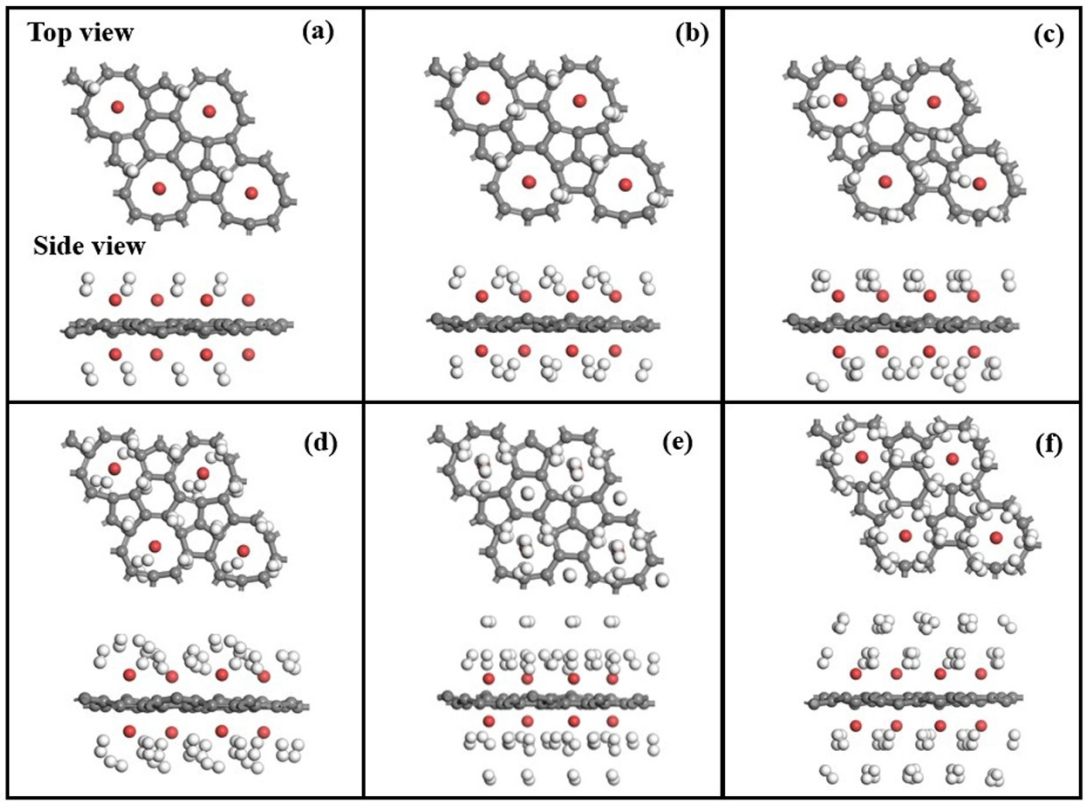

**Fig.5**. The optimized structures of (a) $8H_2$, (b) $16H_2$, (c) $24H_2$, (d) $32H_2$, (e) 40

$H_2$, (f) 48$H_2$ adsorbed on Na-decorated PHE-graphene.

Table 1 The adsorption numbers (n), average adsorption energy ($E_{ad}$), consecutive adsorption energy ($E_c$), desorption temperature of $H_2$ ($T_D$) and the Gravimetric density of 8Li-PHE-graphene and 8Na-PHE-graphene.

| Complex | n | $E_{ad}$(eV/$H_2$) | $E_c$(eV/$H_2$) | $T_D$(K) | Gravimetric density(wt%) |
|---|---|---|---|---|---|
| 8Li-PHE | 8$H_2$ | -0.50 | -0.50 | 639.4 | 2.90 |
|  | 16$H_2$ | -0.38 | -0.26 | 486.0 | 5.60 |
|  | 24$H_2$ | -0.32 | -0.19 | 409.2 | 8.20 |
|  | 32$H_2$ | -0.27 | -0.14 | 345.3 | 10.67 |
|  | 40$H_2$ | -0.23 | -0.07 | 294.1 | 13.00 |
|  | 48$H_2$ | -0.20 | -0.06 | 255.8 | 15.20 |
| 8Na-PHE | 8$H_2$ | -0.32 | -0.32 | 409.2 | 2.40 |
|  | 16$H_2$ | -0.29 | -0.25 | 370.9 | 4.60 |
|  | 24$H_2$ | -0.27 | -0.24 | 345.3 | 6.70 |
|  | 32$H_2$ | -0.23 | -0.12 | 294.1 | 8.80 |
|  | 40$H_2$ | -0.21 | -0.09 | 268.6 | 10.80 |
|  | 48$H_2$ | -0.18 | -0.07 | 230.2 | 12.60 |

In order to reveal the electron transfer of hydrogen molecules during surface adsorption, we calculate the charge density difference (CDD) by Eq. (7). As show in Fig. 6, the yellow and cyan areas figure electron amassing and consumption, respectively, revealing charge transfer between $H_2$ molecules that indicates their polarization into dipoles. This polarization induces a dipole moment between the $H_2$ molecule and the modified structure, resulting in van der Waals forces. In addition, charge exchange occurs between the Li and Na atoms and the modified structure, thus strengthening the ionic bond between the atoms. ELF was used to further study the structure, and the results were obtained as shown in Fig.7. The area near the hydrogen molecule with a value close to 1, showing a high degree of localization, indicating that the hydrogen molecule had formed covalent bonds.

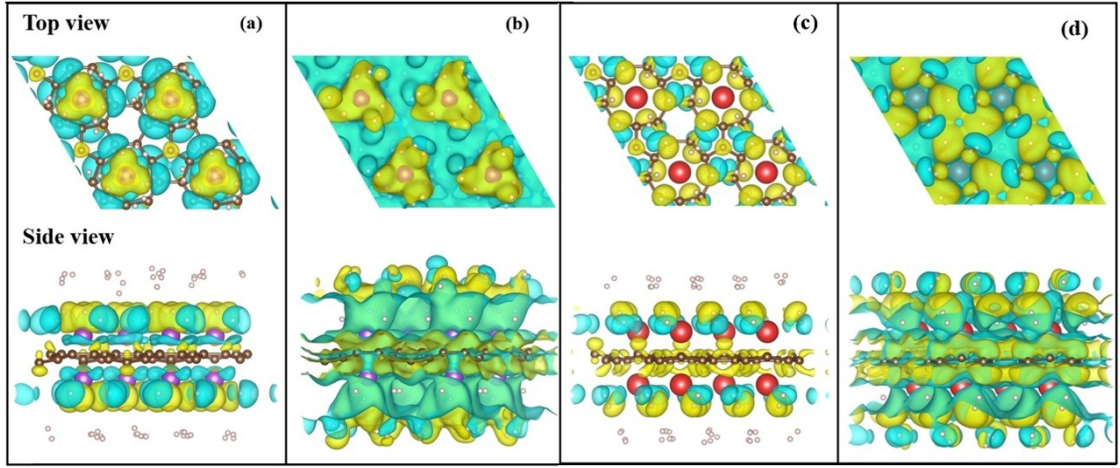

**Fig.6**. Charge density difference of 48H$_2$ absorbed on (a) Li@PHE (Isosurface value is 0.0006 e/ Å$^3$), (b) Li@PHE (Isosurface value is 0.00002 e/ Å$^3$), (c) Na@PHE (Isosurface value is 0.001 e/ Å$^3$), (d) Na@PHE (Isosurface value is 0.0001 e/ Å$^3$).

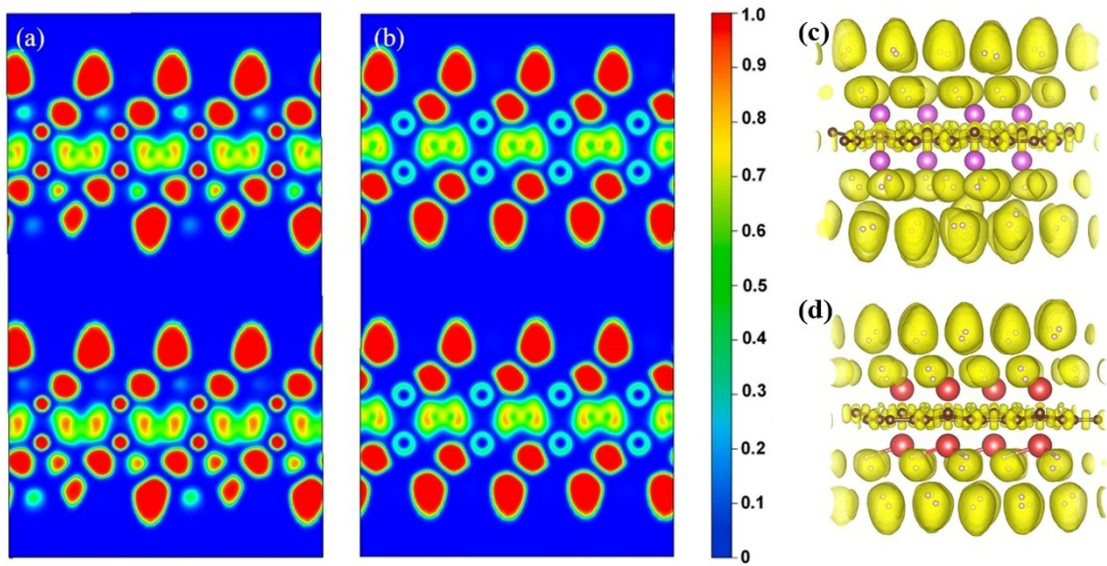

**Fig.7**. The Electron localization function (ELF) of 48H$_2$ absorbed on (a)(c) Li decorated PHE-graphene (b)(d) Na decorated PHE-graphene.

We used MD simulations to evaluate the stability of the structure for hydrogen storage. As illustrated in Fig. 8 during the simulation period, with the temperature undulate around 300 K, the total energy change curve remains relatively stable, manifesting that the H$_2$ adsorbed on the metal-modified PHE-graphene are thermodynamically stable. Based on the aforementioned research, it is demonstrated that Li and Na decorated PHE-graphene is suitable for practical hydrogen storage applications.

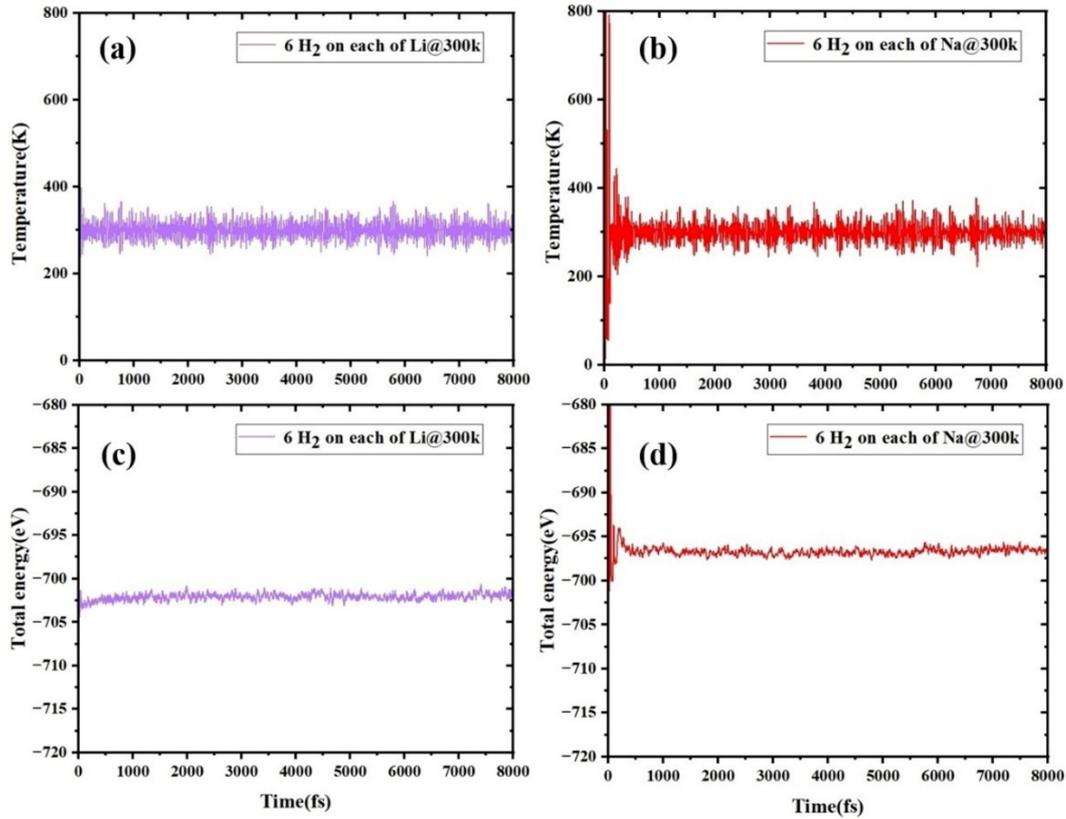

**Fig.8**. Temperature and total energy fluctuation profiles for 6H$_2$ @Li-decorated PHE-graphene ((a) and (c)) and 6H$_2$ @Na-decorated PHE-graphene ((b) and (d)).

### 3.3. The GCMC calculation

We conducted GCMC simulations to assess the hydrogen storage capacity of Li-, Na-modified PHE in practical applications. We constructed the structure illustrated in Fig.9, positioning each hydrogen molecule vertically above the corresponding alkali metal atom. The energy values of hydrogen molecules at different distances from alkali metal atoms were calculated by DFT simulations, leading to the interaction potential energy curves depicted in Fig. 9.

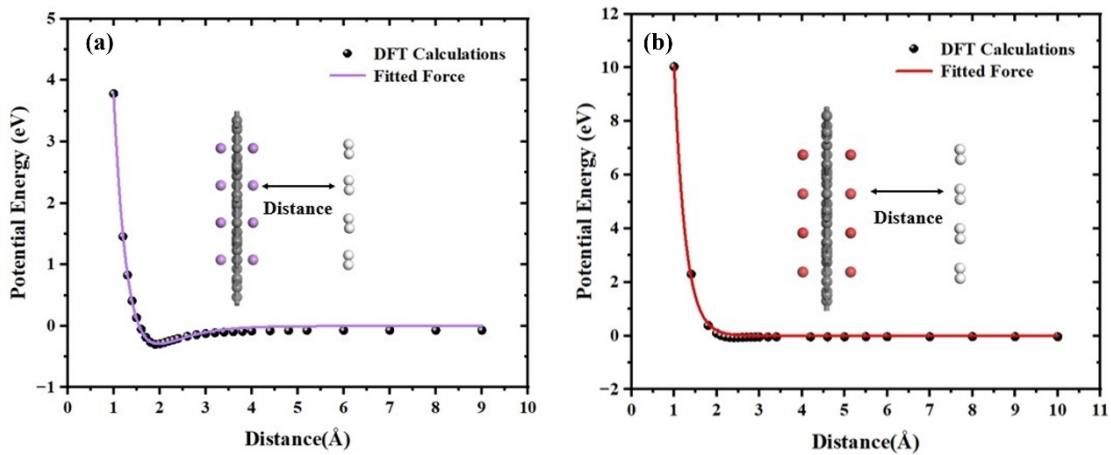

**Fig.9**. Interaction potential energy curves of H$_2$ with (a) Li-decorated PHE, (b) Na-decorated PHE. White: H; gray: C; purple: Li; red: Na.

Using the interaction potential energy curve, we fitted the Morse parameters for Li and Na atoms within the Dreiding force field. The following formula is applied:

$$U(r_{ij}) = D\{exp[\alpha(1 - \frac{r_{ij}}{r_0})] - 2*exp[\frac{\alpha}{2}(1 - \frac{r_{ij}}{r_0})]\} \tag{8}$$

The detailed figures are provided in Table 2, C$_{\_2}$－H$_{\_A}$ has been calculated based on the Dreiding force field [42].

**Table 2** The Morse parameters for Li- and Na- decorated PHE-graphene.

| Term | D (kcal/mol) | R$_0$(Å) | α |
|---|---|---|---|
| Li$_{\_1}$－H$_{\_A}$ | 3.33320 | 2.02214 | 6.58653 |
| Na$_{\_1}$－H$_{\_A}$ | 0.78230 | 2.58343 | 8.79619 |
| C$_{\_2}$－H$_{\_A}$ | 0.06179 | 3.65747 | 5.95610 |

We conducted molecular simulations using GCMC methods with the newly obtained force field parameters at 233 K and 298 K. The isothermal adsorption curves for the hydrogen weight ratio and adsorption enthalpy of Li and Na-modified PHE were obtained with pressure increasing from 1 to 100 bar. The results indicate that the hydrogen weight ratio increases and the enthalpy decreases as pressure rises at constant temperature. As shown in Fig. 10 (a), the Li-modified PHE at 1 bar and 233 K, as well as at 1 bar and 298 K, achieves the 2025 U.S.DOE goal, with the hydrogen storage gravimetric density approaching 10 wt% as the pressure increases. It was previously noted that continuous desorption energies below 0.1 eV indicate negligible adsorption. Referring to the data in Table 1, when 8Li-PHE adsorbs 32 H$_2$ , the consecutive adsorption energy is -0.14 eV, which confirms stable adsorption, and the adsorption weight ratio is 10.67 wt%, aligning well with the reported data. Na-PHE meets the 2025 DOE hydrogen adsorption weight ratio target at 6 bar and 233K, 40bar and 298K. As pressure increases, the hydrogen storage gravimetric density approaches 7.5wt%. Analyzing the data in Table 1, when 32 H$_2$ is adsorbed by 8Na-PHE, the consecutive adsorption energy is -0.12 eV, resulting in an adsorption weight ratio of 8.8 wt%, which is consistent with the findings. The results obtained from DFT calculation and GCMC simulation were basically consistent, validating the accuracy of the data. Previous studies have indicated that the adsorption enthalpy typically ranges from 15 to 25 kJ/mol is favorable for reversible adsorption [43]. As shown in Fig. 10 (b), the adsorption enthalpy at 298K and 233K is essentially around this range. Thus, Li- and Na-modified PHE represents a suitable hydrogen storage material.

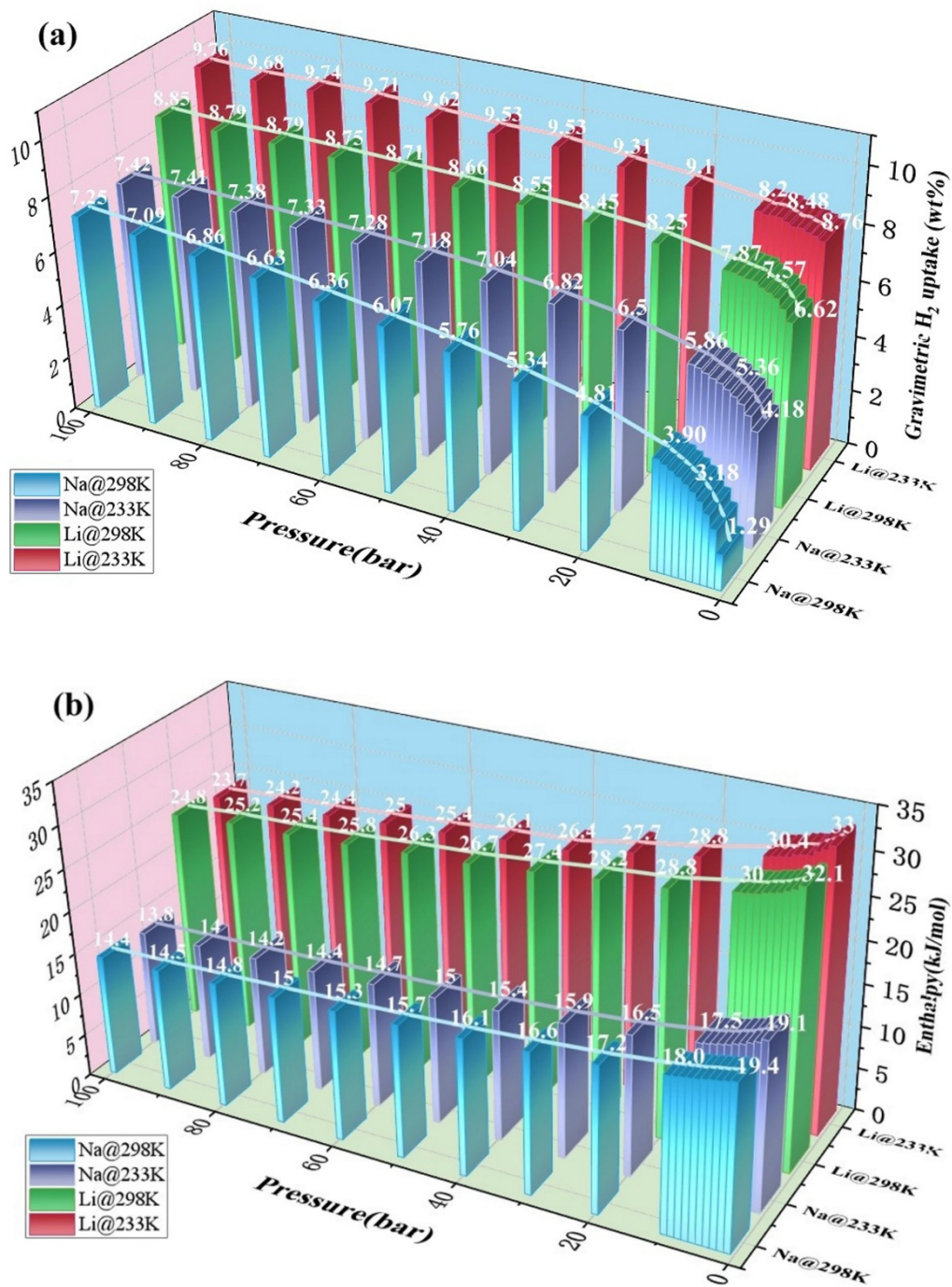

**Fig.10**. The isothermal adsorption curves for (a) the hydrogen weight ratio and (b) adsorption enthalpy of Li- and Na-modified PHE.

## 4.Conclusion

In summary, this study comprehensively probe into the hydrogen adsorption theory of Li- and Na- modified PHE-graphene using DFT calculations. The results indicate that for Li-modified PHE adsorbed six $H_2$ molecules, the average adsorption energy is 0.20 eV, with weight percentage of 15.2wt%. For Na-modified PHE adsorbed six $H_2$ molecules the average adsorption energy is 0.18 eV, with a hydrogen storage weight percentage of 12.6wt%. Furthermore, we continue to conduct GCMC calculation to simulate the maximum hydrogen storage capacity, which corroborates with the DFT calculation results, indicating the reliability of the calculation results. Moreover, electron localization function and charge density difference analyses clearly explain the hydrogen adsorption mechanism. In conclusion, this study highlights that alkali metal modification can enhance hydrogen adsorption, indicating that PHE-graphene could be an excellent material for hydrogen storage.

**CRediT authorship contribution statement**

**Hongyan Ma:** Investigation, Data curation, Writing – original draft. **Qing Wang:** Investigation, Data curation – review & editing. **Huilin Sun:** Data curation, Methodology. **Qingyu Li:** Data curation, Methodology. **Yunhui Wang:** Supervision, Data curation, Funding acquisition, Writing – review& editing. **Zhihong Yang:** Resource, Software, Writing – review& editing. **Huaihong Zhao:** Writing – review& editing. **Huazhong Shu:** Writing – review& editing.

**Declaration of Competing Interest**

The authors declare that they have no known competing financial interests or personal relationships that could have appeared to influence the work reported in this paper.

**Acknowledgments**

This study was supported by National Natural Science Foundation of China (No. 11804169, 11804165), the Natural Science Foundation of Jiangsu Province (No. BK20180741 and 18KJB140011). We also acknowledge the support of Nanjing University of Posts and Telecommuni- cations with No. NY218077.